\documentstyle[prc,aps]{revtex}
\begin{document}
\title{$1/N_c$ Rotational Corrections to $g_A$ in the NJL
Model and Charge Conjugation}
\author{Chr.V.Christov\thanks{Permanent address:
Institute for Nuclear Research and NuclearEnergy, Sofia,
Bulgaria} and
K.Goeke}
\address{Institut f\"ur Theoretische Physik II,
Ruhr-Universit\"at Bochum, D-44780~Bochum, Germany}
\author{P.V.Pobylitsa}
\address{Petersburg Nuclear Physics Institute, Gatchina,
St.Petersburg 188350, Russia}
\maketitle
\begin{abstract}
We show that the $1/N_c$ rotational
corrections to $g_A$, derived using the semiclassical
quantization scheme in the NJL model, possess correct properties under
charge conjugation.
\end{abstract}

Recently, the $1/N_c$ rotational corrections
quantization scheme have been found \cite{Wakamatsu93,Christov94} to
provide a natural solution for the problem of strong
underestimation of the axial-vector coupling constant $g_A$
in the chiral quark soliton (Nambu--Jona-Lasinio) model in leading order.
However, in a very recent paper Schechter and Weigel~\cite{Schechter94}
state that the $1/N_c$ rotational corrections as they are proposed
ref.~\cite{Wakamatsu93} violate the G-parity reflection symmetry and
accordingly they conclude that these
corrections should exactly vanish. Actually they have raised an important
question since the violation of G-parity reflection symmetry would indicate
an inconsistency in the used scheme.

In this brief report we demonstrate that both
the leading order term as well as the $1/N_c$ rotational correction as they
are derived in ref.~\cite{Christov94} possess correct symmetry
properties under charge conjugation.

Similar to ref.~\cite{Schechter94} we choose to work with
the G-parity transformation. Under this transformation
the axial current $\bar\psi \gamma_\mu \gamma_5 \tau^a\psi$ changes sign which
leads to the well-known general relation
\begin{equation}
g_A[{\bar N}] = - g_A[N]\,,
\label{g-A-nucleon-antinucleon}
\end{equation}
between $g_A$ for nucleon $N$ and antinucleon ${\bar N}$.

In the NJL model the formula for
$g_A$, derived~\cite{Christov94} in the semiclassical quantization
scheme, includes leading as well as next to leading order terms:
\begin{equation}
g_A = g_A^{(0)}  +  g_A^{(1)}\,.
\end{equation}
The leading term is given by
\begin{equation}
g_A^{(0)} = -\frac {N_c} 9
\sum\limits_{\mbox{occ. } \textstyle m} \delta_{kb}
\langle m \vert \gamma_0\gamma_5\gamma_k \tau_b \rangle m\vert\,,
\label{g-A-in-chiral-theory-0}
\end{equation}
whereas the next to leading contribution has more complicated
structure\footnote{For simplicity, we use the
non-regularized expressions. However, the regularization will not change our
conclusions. More details about the derivation can be found in
ref.~\cite{Christov94}.}:
\begin{equation}
g_A^{(1)} = \frac {N_c} 9
\sum\limits_{\mbox{occ. } \textstyle m \atop\mbox{non-occ. }\textstyle n}
\frac{i}{2\Theta}  \frac1{\epsilon_n - \epsilon_m} \varepsilon^{akd}
\langle m \vert \gamma_0\gamma_5\gamma_k \tau_d \rangle n \vert
\langle n\vert  \tau_a \rangle  m\vert\,.
\label{g-A-in-chiral-theory-1}\end{equation}
Here, $\Theta$ is the moment of inertia of the soliton. $\epsilon_m$ and
$\rangle m\vert$ are the eigenvalues and eigenstates of the Dirac
Hamiltonian $h(U)$:
\begin{equation}
h(U) = \gamma_0(-i\gamma^k\partial_k + MU^{\gamma_5}+m_0) \quad
\mbox{and}\quad
h(U)ranglem(U)\vert = E_m(U) \rangle m(U)\vert
\end{equation}
In the case of nucleon (baryon number one solution) the
occupied states inclide the Dirac sea and the valence level.

The meson field $U$ has hedgehog symmetry
\begin{equation}
U(x)= \mbox{e}^{i P(|{\vec x}|) (x^a \tau^a)/|{\vec x}|}\,.
\label{U(x)}
\end{equation}
which survives under G-parity transformation:
\begin{equation}
G U(x) G^{-1} = U^\dagger(x).
\end{equation}
This means that antinucleon is described by the hedgehog soliton with meson
field $U^\dagger(x)$. Accordingly the relation \ref{g-A-nucleon-antinucleon}
can be rewritten as
\begin{equation}
 g_A[U^\dagger] = - g_A[U]\,.
\label{g-A-nucleon-antinucleon1}
\end{equation}
It should be noted that both charge conjugation and G-parity transformation
can be used to obtain antinucleon state. However, the
G-parity transformation is more convenient since it explicitly preserves
both the isospin and the hedgehog ansatz whereas in the case of the charge
conjugation one should use additionally the invariance of the hedgehog
solution \ref{U(x)} under SU(2)-isorotation. It should be also stressed that
both transformations lead to one and the same result
\ref{g-A-nucleon-antinucleon1}.

Using the identity
\begin{equation}
h(U^\dagger) = - (\gamma_0\gamma_5) h(U) (\gamma_0\gamma_5)^{-1}\,,
\end{equation}
it easy to see that following relations are valid:
\begin{equation}
\rangle{m(U^\dagger)}\vert = \gamma_0\gamma_5 \rangle{m(U)}\vert\qquad
\mbox{and}\qquad \epsilon_m(U^\dagger) = - \epsilon_m(U)\,.
\label{states-transformation}
\end{equation}

Let us start from the leading contribution $g_A^{(0)}$
\ref{g-A-in-chiral-theory-0}.
First, from \ref{states-transformation} we have
\begin{equation}
 {m(U^\dagger)}  \gamma_0\gamma_5\gamma_k \tau_b \rangle m(U^\dagger)\vert
= \langle {m(U)}\vert  \gamma_0\gamma_5\gamma_k \tau_b \rangle
m(U)\vert \,. \end{equation}
Second, using the identity
\begin{equation}
\sum\limits_{\mbox{occ. } \textstyle m(U)} \hskip-0.5cm\delta_{kb}
\langle m(U)\vert   \gamma_4\gamma_5\gamma_k \tau_b  \rangle m(U)\vert
= -\hskip-1cm \sum\limits_{\mbox{non-occ.}\textstyle m(U)}\hskip-0.5cm
\delta_{kb} \langle m(U)\vert  \gamma_4\gamma_5\gamma_k \tau_b \rangle
m(U)\vert \,, \end{equation}
and also the fact that the occupied states of $h(U)$ correspond to non-occupied
states of $h(U^\dagger)$ we find in agreement with
\ref{g-A-nucleon-antinucleon1}:
\begin{equation}
 g_A^{(0)}[U^\dagger] = - g_A^{(0)}[U]\,.
\end{equation}

For the rotational correction $g_A^{(1)}$
in the case of antinucleon using again relations \ref{states-transformation}
we have
\begin{eqnarray}
&&\sum\limits_{\mbox{occ. }
\textstyle m(U^\dagger) \atop\mbox{non-occ. }\textstyle n(U^\dagger)}
 \frac1{\epsilon_n(U^\dagger) - \epsilon_m(U^\dagger)}
\langle m(U^\dagger)\vert
\gamma_0\gamma_5\gamma_k \tau_d \rangle n(U^\dagger)\vert
 \langle n(U^\dagger)\vert    \tau_a \rangle m(U^\dagger)\vert
\nonumber\\
&&=-\hskip-0.8cm\sum\limits_{\mbox{non-occ. }
\textstyle n(U) \atop\mbox{occ. }\textstyle m(U)}
 \frac 1{ \epsilon_n(U)- \epsilon_m(U)}
\langle  m(U)\vert
\gamma_0\gamma_5\gamma_k \tau_d \rangle n(U)\vert \langle n(U)\vert \tau_a
\rangle m(U)\vert \,. \end{eqnarray}
Further we make use of the properties of the matrix elements
under the interchange $ m \longleftrightarrow n $
\begin{equation}
\langle m\vert
 \gamma_0\gamma_5\gamma_k \tau_d \rangle n\vert  \langle n\vert   \tau_a
\rangle m\vert = - \langle n\vert
\gamma_0\gamma_5\gamma_k \tau_d \rangle m\vert  \langle m\vert   \tau_a
\rangle n\vert\,, \end{equation}
to get the final result:
\begin{equation}
 g_A^{(1)}[U^\dagger] = - g_A^{(1)}[U]\,.
\end{equation}
One concludes that the next to leading order term shows the same G-parity
properties
as the leading one and both agree with  \ref{g-A-nucleon-antinucleon1}.

Actually the expression \ref{g-A-in-chiral-theory-1}, studied
above, are derived in a formalism based on the path integral approach as it
is presented in ref.~\cite{Christov94}. To be more particular, in the
derivation of the $1/N_c$ rotational corrections to $g_A$ using
semiclassical quantization not-commuting collective operators appear
and the final result depends on the ordering of these operators. Our scheme
involves time-ordering of collective operators which follows from the
path integral without any ambiguity, applicable of course to both nucleon
and antinucleon.  In this respect it is not surprising that G-parity
properties  are accurately described. In contrast to us, the scheme of
Wakamatsu and Watabe~\cite{Wakamatsu93} uses a different ordering
whose theoretical origin is not justified. Hence their formulae are
different from the ones of ref.~\cite{Christov94} and in particular, as it
has been shown in ref.~\cite{Schechter94} it violates the G-parity
reflection symmetry.

{\bf Acknowledgement:}

The project has been partially supported
by the VW Stiftung and DFG.

\end{document}